\documentclass[reprint,amsmath,amssymb,aps,prb,showpacs,floatfix,superscriptaddress]{revtex4-1}

\usepackage{times,color,graphicx}
\usepackage[colorlinks=true,urlcolor=blue,linkcolor=blue,
            citecolor=blue]{hyperref}

\newcommand{\bibtitle}[1]{\textit{#1},}

\begin{document}

\title{Stochastic sampling of quadrature grids for the
       evaluation of vibrational expectation values}

\author{Pablo L\'opez R\'ios}
  \email{pl275@cam.ac.uk}
  \affiliation{Max-Planck Institute for Solid State Research,
               Heisenbergstra{\ss}e 1, 70569 Stuttgart, Germany}
  \affiliation{Theory of Condensed Matter Group, Cavendish Laboratory,
               J.\ J.\ Thomson Avenue, Cambridge CB3 0HE, UK}

\author{Bartomeu Monserrat}
  \affiliation{Theory of Condensed Matter Group, Cavendish Laboratory,
               J.\ J.\ Thomson Avenue, Cambridge CB3 0HE, UK}
  \affiliation{Department of Physics and Astronomy,
               Rutgers University, Piscataway,
               New Jersey 08854-8019, USA}

\author{Richard J.\ Needs}
  \affiliation{Theory of Condensed Matter Group, Cavendish Laboratory,
               J.\ J.\ Thomson Avenue, Cambridge CB3 0HE, UK}

\date{\today}

\begin{abstract}
  The \textit{thermal lines} method for the evaluation of vibrational
  expectation values of electronic observables
  [\href{https://doi.org/10.1103/PhysRevB.93.014302}{B. Monserrat,
  Phys.\ Rev.\ B \textbf{93}, 014302 (2016)}] was recently proposed as
  a physically motivated approximation offering balance between the
  accuracy of direct Monte Carlo integration and the low computational
  cost of using local quadratic approximations.
  In this paper we reformulate thermal lines as a stochastic
  implementation of quadrature grid integration, analyze the
  analytical form of its bias, and extend the method to multiple point
  quadrature grids applicable to any factorizable harmonic or
  anharmonic nuclear wave function.
  The bias incurred by thermal lines is found to depend on the local
  form of the expectation value, and we demonstrate that the use of
  finer quadrature grids along selected modes can eliminate this bias,
  while still offering a $\sim 30\%$ lower computational cost than
  direct Monte Carlo integration in our tests.
\end{abstract}


\maketitle

\section{Introduction}
\label{sec:introduction}

First-principles studies of solid state systems typically use the
Born-Oppenheimer approximation \cite{born_molecules_1929} to simplify
the task of solving the Schr\"odinger equation of the system by
separating the electronic and nuclear degrees of freedom.
Density functional theory (DFT) \cite{jones_dft_2015} is the
\textit{de facto} standard method for solving the electronic problem
in crystalline systems, allowing the numerical evaluation of a large
number of relevant properties of materials at a relatively low
computational cost.
The approximate nature of density functionals proves problematic in
some cases, and more accurate, computationally costly methods such as
the $GW$ approximation \cite{Hedin_GW_1965,Aryasetiawan_GW_1998} or
quantum Monte Carlo \cite{ceperley_qmc_1980, foulkes_qmc_2001} can be
used instead for the evaluation of ground-state observables.

Many electronic properties can usually be evaluated accurately
within the static lattice approximation, in which the electronic
problem is solved neglecting the effects of nuclear motion entirely.
However, zero-point quantum corrections play a crucial role in systems
containing light elements, and the description of thermal effects
requires the inclusion of vibrations in all systems.
The harmonic approximation \cite{wallace_thermodynamics_1971,
born_dynamical_1956, maradudin_harmonic_1971} enables the inclusion of
nuclear motion for the calculation of the total energy at small
computational cost.
Although this approach is very accurate for most systems, there are
notable exceptions including systems with light atoms,
\cite{errea_h3s_2016} those involving weak bonding between atoms,
\cite{fultz_scf3_2011} those at high temperature,
\cite{Zhao_SnSe_2014} or those near structural phase transitions.
\cite{trail_tio2_2017}
Anharmonic approximations \cite{hooton_anharmonic_1958,
souvatzis_stabilization_2008, hellman_anharmonic_2011,
antolin_free_energy_2012, monserrat_anharmonic_2013,
errea_anharmonicity_2013} exist that are able to overcome these
limitations at an additional computational cost.

Vibrational corrections to electronic properties other than the energy
can be calculated as the expectation value of the property of interest
with respect to the vibrational harmonic or anharmonic nuclear wave
function.
Literature examples of such calculations include thermal averages of
the electronic band structure, \cite{allen_band_structure_1981,
allen_gap_1981, giustino_diamond_2010} the dielectric function,
\cite{williams_1951, lax_1952, patrick_epsilon_2014,
zacharias_epsilon_2015} the chemical shielding tensor,
\cite{rossano_nmr_2005, dumez_nmr_2009, monserrat_nmr_2014,
dracinsky_nmr_2014, nemausat_nmr_2015} the x-ray absorption near-edge
structure, \cite{nemausat_nmr_2015} or the contact hyperfine
interaction. \cite{porter_muon_1999, moller_muon_2013}
These vibrational averages have traditionally been computed using
either a quadratic expansion of the electronic observable around
its equilibrium value, or Monte Carlo integration.
%
%
The former approach is computationally advantageous, but neglects
multi-phonon terms in the electron-phonon interaction.
The latter approach is computationally costly, but multi-phonon terms
are correctly captured.
The recent demonstration that multi-phonon terms can be important in a
number of cases \cite{monserrat_helium_2014, monserrat_giant_2015,
patrick_cssni3_2015, saidi_mapbi3_2016} has motivated the development
of the thermal lines method, \cite{monserrat_TL_2016} which can
partially capture multi-phonon terms at a smaller computational cost
than Monte Carlo integration.

The thermal lines method was proposed as a physically-motivated
approximation to the evaluation of vibrational expectation values of
electronic observables, offering faster statistical convergence than
Monte Carlo integration and better accuracy than the quadratic
approximation, but its bias has not yet been quantified.
In the present paper we analyze the bias of the thermal lines method,
and use this analysis to propose an integration method based on the
stochastic sampling of quadrature grids constructed from the
vibrational wave function.
The bias in our proposed method is controlled by the number of grid
points, reducing to thermal lines for a two-point grid and to Monte
Carlo integration in the infinite-point limit.
Our results indicate that thermal lines (two-point grids) are capable
of accurately incorporating the effects of multi-phonon terms in most
calculations of vibrational expectation values.
However, we find that the bias incurred by thermal lines is not
negligible for observables with a strong non-quadratic behavior, and
multiple point grids are required in order to obtain accurate
expectation values in these cases.

The rest of this paper is structured as follows.
In section \ref{sec:vibrations} we provide the theoretical background
and describe the relevant methods for the evaluation of vibrational
expectation values.
The bias in the thermal lines method is analyzed in section
\ref{sec:TL_analysis}.
In section \ref{sec:1d_quadrature} we formulate and test
one-dimensional quadrature grids, and in section
\ref{sec:mc_quadrature} we discuss the use of these grids in
multi-dimensional Monte Carlo sampling for realistic examples of
vibrational averages of electronic observables.
Finally, we summarize our findings in section \ref{sec:conclusions}.
Hartree atomic units ($\hbar=|e|=m_{\rm e}=4\pi\epsilon_0=1$)
are used throughout.

\section{Vibrational calculations}
\label{sec:vibrations}

In computer simulations, crystalline solids are represented by
supercells containing $N$ nuclei and $N_e$ electrons subject to
periodic boundary conditions.
Within the Born-Oppenheimer approximation, the electronic Hamiltonian
${\hat H}_{\rm el}({\bf R})$ parametrically depends on the nuclear
positions ${\bf R}$, and the resulting electronic energy
$E_{\rm el}({\bf R})$, in which we include nucleus-nucleus
interactions, acts as an external potential in the vibrational
Hamiltonian,
\begin{equation}
  \label{eq:hamiltonian_vib_r}
  {\hat H}_{\rm vib}({\bf R}) =
    - \sum_{\alpha=1}^N \frac 1 {2m_\alpha} \nabla_{\alpha}^2
    + E_{\rm el}({\bf R}) \;.
\end{equation}
Without loss of generality we will assume that $E_{\rm el}({\bf R}_0)
= 0$, where ${\bf R}_0$ is the equilibrium nuclear configuration.
The nuclear motion problem has three translational degrees of freedom
which can be eliminated, and $n=3N-3$ vibrational degrees of freedom.

It is convenient to work in terms of normal-mode coordinates ${\bf u}
= \{u_i\}_{i=1}^n$, which are the real-valued linear combinations of
nuclear displacements under which the dynamical matrix is diagonal.
\cite{wallace_thermodynamics_1971, born_dynamical_1956,
maradudin_harmonic_1971}
We replace the phonon branch $\nu$ and wave vector ${\bf q}$, which
are the standard normal mode quantum numbers, by a single index $i$
for notational convenience.
Note that ${\bf u}={\bf 0}$ at the equilibrium nuclear configuration
${\bf R}_0$.
In these coordinates, Eq.\@ \ref{eq:hamiltonian_vib_r} reads
\begin{equation}
  \label{eq:hamiltonian_vib_q}
  {\hat H}_{\rm vib}({\bf u}) =
    - \frac 1 2 \sum_{i=1}^n \frac{\partial^2}{\partial u_i^2}
    + E_{\rm el}({\bf u}) \;.
\end{equation}
The ground-state vibrational wave function $\Phi({\bf u})$ determines
the zero-temperature quantum vibrational properties of the system, and
can be obtained by solving the vibrational Schr\"odinger equation,
${\hat H}_{\rm vib}({\bf u})\Phi({\bf u}) = E_{\rm vib} \Phi({\bf
u})$.

Under the harmonic approximation, the electronic energy is assumed to
take the form $E_{\rm el}({\bf u}) = \sum_i \frac 1 2 \omega_i u_i^2$,
where the harmonic frequency of the $i$th normal mode $\omega_i$ can
be obtained by differentiating the electronic energy, either using
finite differences \cite{fd_kunc_1982, nondiagonal_lloydwilliams_2015}
or methods such as density functional perturbation theory,
\cite{baroni_dfpt_2001} with respect to $u_i$.
The harmonic approximation results in a wave function of the form
$\Phi^{\rm (har)}({\bf u}) = \prod_{i=1}^n \phi^{(\rm har)}_i(u_i)$,
where
\begin{equation}
  \label{eq:harmonic_wfn}
  \phi^{\rm (har)}_i(u_i) = \left( \frac {\omega_i} \pi\right)^{1/4}
  \exp\left(-\frac 1 2 \omega_i u_i^2\right) \;.
\end{equation}
Various methods for dealing with anharmonic vibrational Hamiltonians
exist which lift the restriction on the form of the electronic energy,
\cite{souvatzis_stabilization_2008, hellman_anharmonic_2011,
antolin_free_energy_2012, monserrat_anharmonic_2013,
errea_anharmonicity_2013} but all these methods assume that the wave
function can be factorized as a product of single-mode functions,
$\Phi({\bf u}) = \prod_{i=1}^n \phi_i(u_i)$.
In our analysis we assume the use of a factorizable vibrational wave
function.

We focus our present discussion on the zero-temperature limit for
simplicity.
Thermal effects can be trivially including by replacing the
ground-state vibrational density $\Phi^2$ with $\frac 1 {\cal Z}
\sum_{\bf s}\Phi_{\bf s}^2 e^{\beta E_{\bf s}}$, where $\beta =
({k_{\rm B} T})^{-1}$ is the inverse temperature, $\bf s$ identifies
the vibrational excited state corresponding to wave function
$\Phi_{\bf s}$ and energy $E_{\bf s}$, and ${\cal Z} = \sum_{\bf s}
e^{\beta E_{\bf s}}$ is the partition function of the system.
We note that the ``mean thermal line'' approach for evaluating
finite-temperature expectation values \cite{monserrat_TL_2016} remains
applicable to our proposed method.

\subsection{Vibrational expectation values}

Knowledge of $\Phi({\bf u})$ enables the evaluation of the expectation
value of an electronic observable $\hat A$ with respect to the
vibrational density as
\begin{equation}
  \label{eq:expval_def}
  \langle A\rangle_{\Phi^2} =
    \int \Phi^2({\bf u}) A({\bf u}) \, {\rm d}{\bf u} \;,
\end{equation}
where $A({\bf u}) = \Phi^{-1}({\bf u}) {\hat A} \Phi({\bf u})$ is the
local value of the observable at nuclear configuration $\bf u$.
Under the assumption that $A({\bf u})$ is a smooth function of
its arguments, it is useful to express it as a power expansion,
\begin{eqnarray}
  \label{eq:a_expand}
  A({\bf u})
  &=& A({\bf 0}) + \sum_i^n a_i^{(1)} u_i
      + \sum_{i\leq j}^n a_{ij}^{(2)} u_i u_j
      \nonumber \\
  &+& \sum_{i\leq j\leq k}^n a_{ijk}^{(3)} u_i u_j u_k
      + \ldots \;,
\end{eqnarray}
where $a_i^{(1)}$, $a_{ij}^{(2)}$, $a_{ijk}^{(3)}$, etc., are linear
expansion coefficients.
It should be noted that the approximations involved in solving the
vibrational Schr\"odinger equation are distinct from those applied to
$A({\bf u})$; namely, the use of the harmonic wave function does not
imply that expectation values are assumed to be quadratic functions of
$\bf u$ (or \textit{vice versa}), and the use of a factorizable wave
function without explicit phonon-phonon correlations does not imply
neglecting multi-mode contributions from Eq.\@ \ref{eq:a_expand} (or
\textit{vice versa}). \cite{antonius_anh_2015}

\subsubsection{The quadratic approximation}

If $\Phi({\bf u})$ is symmetric, such as in the harmonic
approximation, substituting Eq.\ \ref{eq:a_expand} into Eq.\@
\ref{eq:expval_def} results in the cancellation of all contributions
involving odd powers of a normal-mode coordinate,
\begin{eqnarray}
  \nonumber
  \langle A \rangle_{\Phi^2}
    &=& A({\bf 0})
        + \sum_i^n a_{ii}^{(2)}
            \langle
              u_i^2
            \rangle_{\Phi^2}
        + \sum_{i\leq j}^n a_{iijj}^{(4)}
            \langle
              u_i^2 u_j^2
            \rangle_{\Phi^2} \\
  \label{eq:expval_quad_plus}
    &+& \sum_{i\leq j\leq k}^n a_{iijjkk}^{(6)}
            \langle
              u_i^2 u_j^2 u_k^2
            \rangle_{\Phi^2}
        + \ldots \;,
\end{eqnarray}
and since $\Phi^2$ is factorizable, Eq.\ \ref{eq:expval_quad_plus}
reduces to a linear combination of products of one-dimensional
integrals.
Neglecting fourth- and higher-order contributions to Eq.\@
\ref{eq:expval_quad_plus} yields the \textit{quadratic
approximation},
\begin{equation}
  \label{eq:expval_quad}
  \langle A \rangle_{\Phi^2} \approx
    A({\bf 0}) + \sum_i^n a_{ii}^{(2)}
      \langle u_i^2 \rangle_{\phi_i^2} \;,
\end{equation}
where $a_{ii}^{(2)}=\frac 1 2 \left(
\frac{\partial^2 A}{\partial u_i^2}\right)_{{\bf u}={\bf 0}}$,
and the one-dimensional integrals $\langle u_i^2 \rangle_{\phi_i^2}$
can easily be calculated; in the harmonic approximation,
$\langle u_i^2 \rangle_{\phi_i^2} = (2\omega_i)^{-1}$.

\subsubsection{Direct Monte Carlo integration}

An unbiased estimate of the integral of Eq.\ \ref{eq:expval_def} can
be evaluated using Monte Carlo integration,
\begin{equation}
  \label{eq:expval_mc}
  \langle A \rangle_{\Phi^2} \approx
  \frac 1 M \sum_{m=1}^M A({\bf u}_m) \;,
\end{equation}
where $\{{\bf u}_m\}$ are $M$ random vectors of normal-mode
coordinates distributed according to $\Phi^2({\bf u})$.
This method, which we refer to as \textit{direct Monte Carlo} in what
follows, is exact in the limit $M\to\infty$, and requires no
assumptions about the form of $A({\bf u})$ or $\Phi({\bf u})$.
However, if the wave function is symmetric about ${\bf u}={\bf 0}$,
$\Phi({\bf u})=\Phi(-{\bf u})$, it is advantageous to accumulate
samples in \{${\bf u}$, $-{\bf u}$\} pairs in order to exactly
remove odd-order contributions,
\begin{eqnarray}
  \nonumber
  A^*({\bf u}) &=&
  \frac 1 2 [A({\bf u}) + A(-{\bf u})] \\
  &=& A({\bf 0}) + \sum_{i\leq j}^n a_{ij}^{(2)} u_i u_j
      \nonumber \\
  \label{eq:a_expand_symm}
  &+&
 \sum_{i\leq j\leq k \leq l}^n a_{ijkl}^{(4)} u_i u_j u_k u_l
      + \ldots \;,
\end{eqnarray}
resulting in reduced random noise and faster statistical convergence,
while giving the correct expectation value since it is trivial
that $\langle A({\bf u})\rangle_{\Phi^2} = \langle A^*({\bf
u})\rangle_{\Phi^2}$.
This sampling strategy, which we refer to as \textit{symmetrized
sampling}, was used in Ref.\@ \onlinecite{monserrat_TL_2016} for
the ``TL$_2$'' variant of the thermal lines method, but it can be
used to accelerate any Monte Carlo evaluation of the expectation value
of $\hat A$ with a symmetric wave function.

\subsubsection{Thermal lines}

Inspired by the mean value theorem for integrals, the thermal lines
method \cite{monserrat_TL_2016} postulates that a
good approximation to the expectation value of $\hat A$ for a
symmetric wave function is given by
\begin{equation}
  \label{eq:TL_mc}
  \langle A \rangle_{\Phi^2}\approx
  \frac 1 M \sum_{m=1}^M A[{\bf u}_{\rm TL}({\bf S}_m)] \;,
\end{equation}
where $\{{\bf S}_m\}$ are $M$ random $n$-dimensional vectors each of
whose components takes the values $+1$ and $-1$ with equal
probability, and ${\bf u}_{\rm TL}({\bf S})$ is a vector whose $i$th
component is $S_i U_i$, where $U_i =
\sqrt{\langle u_i^2 \rangle_{\Phi^2}}$.

The thermal lines method defines ``special'' points at which to sample
the integrand, much in the spirit of special $k$-point methods for
integrals over the Brillouin zone of a crystal,
\cite{baldereschi_integration_1973, rajagopal_specialk_1994} or
quadrature grids for integrals over the surface of a sphere,
\cite{lebedev_quadrature_1976, reeger_quadrature_2015} and uses Monte
Carlo sampling of these points to efficiently deal with the high
dimensionality of the integration volume.
Although the effectiveness of thermal lines for the calculation of
vibrational averages was demonstrated in Ref.\@
\onlinecite{monserrat_TL_2016}, there was no formal analysis of the
bias incurred by replacing Eq.\ \ref{eq:expval_mc} by Eq.\@
\ref{eq:TL_mc}.
We present such analysis in section \ref{sec:TL_analysis}.

We note that Monte Carlo sampling over thermal lines has been used to
study the effects of electron-phonon coupling on the temperature
dependence of the band gaps of a number of semiconductors within the
$GW$ approximation \cite{Monserrat_GW_2016} and to study
phonon-assisted optical absorption in BaSnO$_3$ using a hybrid DFT
functional. \cite{Monserrat_BaSnO3_2017}
It has also been shown that, in the thermodynamic limit $n\to\infty$,
a single thermal line delivers the exact thermal average $\langle
A\rangle_{\Phi^2}$ without the need of Monte Carlo sampling.
This thermal line is such that the sign $S_i$ alternates between $+1$
and $-1$ when the coordinates $u_i$ are ordered by increasing value of
their associated harmonic frequencies $\omega_i$.
\cite{zacharias_one-shot_2016}
Finally, we note that other methods in the same spirit as thermal
lines have been used, for example in the study of superconducting
hydrogen sulfides. \cite{Komelj_H3S_2015}

\section{Analysis of thermal lines}
\label{sec:TL_analysis}

The distribution of normal-mode configurations $\bf u$ sampled in
Eq.\ \ref{eq:TL_mc} can be identified with the following probability
density function,
\begin{equation}
  \label{eq:TL_wfn}
  \Phi^2_{\rm TL}({\bf u}) =
    \frac 1 {2^n}
    \prod_{i=1}^n
      \left[
        \delta(u_i-U_i) +
        \delta(u_i+U_i)
      \right] \;,
\end{equation}
where $\Phi_{\rm TL}$ is the ``thermal lines wave function''.
At ${\bf u}_{\rm TL}= \left\{S_i U_i \right\}$,
Eq.\ \ref{eq:a_expand} becomes
\begin{eqnarray}
  \nonumber
  A({\bf u}_{\rm TL})
  &=& A({\bf 0})
      + \sum_{i=1}^n a_{i}^{(1)} S_i U_i
      + \sum_{i\leq j}^n a_{ij}^{(2)} S_i S_j U_i U_j \\
  \nonumber
  &+& \sum_{i\leq j\leq k}^n a_{ijk}^{(3)} S_i S_j S_k U_i U_j U_k \\
  \label{eq:a_expand_TL}
  &+& \sum_{i\leq j\leq k\leq l}^n a_{ijkl}^{(4)} S_i S_j S_k S_l
                                     U_i U_j U_k U_l + \ldots \;.
\end{eqnarray}
The expectation value of products of powers of $S$ for different
modes factorizes into products of single-mode expectation values,
e.g., $\langle S_i^\beta S_j^\gamma\rangle =
\langle S_i^\beta\rangle \langle S_j^\gamma\rangle$
for $i\neq j$, and each of these is zero if the exponent is odd and
unity if the exponent is even.
The expectation value of $\hat A$ under the thermal lines wave
function is therefore
\begin{eqnarray}
  \nonumber
  \langle A\rangle_{\Phi_{\rm TL}^2}
  &=& A({\bf 0})
      + \sum_i^n a_{ii}^{(2)}
          U_i^2
      + \sum_{i\leq j}^n a_{iijj}^{(4)}
          U_i^2 U_j^2 \\
  \label{eq:expval_TL}
  &+& \sum_{i\leq j\leq k}^n a_{iijjkk}^{(6)}
          U_i^2 U_j^2 U_k^2
      + \ldots \;,
\end{eqnarray}
which agrees with the quadratic approximation to second order,
but includes higher-order multi-mode contributions.
Subtracting Eq.\@ \ref{eq:expval_TL} from Eq.\@
\ref{eq:expval_quad_plus} and substituting $U_i^2 =
\langle u_i^2\rangle_{\Phi^2}$, the bias in the thermal lines
expectation value is
\begin{eqnarray}
  \nonumber
  \langle A \rangle_{\Phi^2}
    - \langle A\rangle_{\Phi_{\rm TL}^2}
  &=& \sum_i^n a_{iiii}^{(4)}
               \left(
                  \langle u_i^4 \rangle_{\Phi^2} -
                  \langle u_i^2 \rangle_{\Phi^2}^2
               \right) \\
  \nonumber
  &+& \sum_i^n a_{iiiiii}^{(6)}
               \left(
                  \langle u_i^6 \rangle_{\Phi^2} -
                  \langle u_i^2 \rangle_{\Phi^2}^3
               \right) \\
  \nonumber
  &+& \sum_{i<j}^n a_{iiiijj}^{(6)}
                   \left(
                      \langle u_i^4 \rangle_{\Phi^2} -
                      \langle u_i^2 \rangle_{\Phi^2}^2
                   \right)
                   \langle u_j^2 \rangle_{\Phi^2} \\
  \nonumber
  &+& \sum_{i<j}^n a_{iijjjj}^{(6)}
                   \langle u_i^2 \rangle_{\Phi^2}
                   \left(
                      \langle u_j^4 \rangle_{\Phi^2} -
                      \langle u_j^2 \rangle_{\Phi^2}^2
                   \right) \\
  \label{eq:expval_TL_error}
  &+& {\cal O}(u^8) \;.
\end{eqnarray}
It should be noted that the bias in the quadratic approximation
includes all fourth- and higher-order terms in Eq.\@
\ref{eq:expval_quad_plus}, while Eq.\@ \ref{eq:expval_TL_error} shows
that the bias in thermal lines arises solely from terms of fourth
or higher order in which all index values appear an even number of
times.
Therefore the bias in thermal lines expectation values can typically
be expected to be smaller than that for the quadratic approximation.

The variance of the values of $A({\bf u})$ encountered during Monte
Carlo integration determines the statistical uncertainty of the
result.
The sample variance ${\rm var}[A] = \left\langle \left(A-\langle
A\rangle\right)^2 \right\rangle$ associated with the $\Phi^2$
distribution is
\begin{eqnarray}
\nonumber
{\rm var}_{\Phi^2}[A] &=&
  \sum_{i=1}^n {a_i^{(1)}}^2 \langle u_i^2\rangle_{\Phi^2}
  \\ \nonumber &+&
  \sum_{i=1}^n {a_{ii}^{(2)}}^2 \left( \langle u_i^4\rangle_{\Phi^2} -
    \langle u_i^2\rangle_{\Phi^2}^2 \right)
  \\ \nonumber &+&
  \sum_{i<j}^n {a_{ij}^{(2)}}^2 \langle u_i^2\rangle_{\Phi^2}
    \langle u_j^2\rangle_{\Phi^2}
  \\ \nonumber &+&
  \sum_{i=1}^n a_i^{(1)} a_{iii}^{(3)}
    \langle u_i^4\rangle_{\Phi^2}
  \\ \nonumber &+&
  \sum_{i<j}^n \left( a_i^{(1)} a_{iij}^{(3)} +
    a_j^{(1)} a_{ijj}^{(3)} \right)
    \langle u_i^2\rangle_{\Phi^2}^2
    \langle u_j^2\rangle_{\Phi^2}^2
  \\ \label{eq:expval_variance} &+&
    {\cal O}(u^6) \;.
\end{eqnarray}
The expression for ${\rm var}_{\Phi_{\rm TL}^2}[A]$ is similar to Eq.\@
\ref{eq:expval_variance}, but replacing $\langle
u_i^4\rangle_{\Phi^2}$ with $\langle u_i^2\rangle_{\Phi^2}^2$, which
eliminates the second term.
The leading order contribution to both ${\rm var}_{\Phi^2}[A]$ and
${\rm var}_{\Phi_{\rm TL}^2}[A]$ is thus due to the asymmetry of the
expectation value along individual modes.
We note that there exist observables for which the thermal lines
method gives a greater variance than direct Monte Carlo integration.
For example, for a one-dimensional harmonic wave function of frequency
$\omega=1$ the function $A(u) = -3u +u^2 +u^3$ is sampled with greater
variance with unsymmetrized-sampling thermal lines than with
unsymmetrized-sampling direct Monte Carlo.
However, in the absence of asymmetries, the cancellation of the
second term in Eq.\@ \ref{eq:expval_variance} implies that
${\rm var}_{\Phi_{\rm TL}^2}[A] < {\rm var}_{\Phi^2}[A]$ to leading
order.

The variance of $A^*$ can be obtained from Eq.\@
\ref{eq:expval_variance} by zeroing the $a$ coefficients of odd-order
terms, so only the second and third terms survive in
${\rm var}_{\Phi^2}[A^*]$, and only the third term survives in
${\rm var}_{\Phi_{\rm TL}^2}[A^*]$.
Thus, the variance from symmetrized-sampling thermal lines is
identically zero in one dimension, and in multiple dimensions the
sample variance arises solely from multi-mode contributions.
By contrast, ${\rm var}_{\Phi^2}[A^*]$ contains single-mode
contributions, hence ${\rm var}_{\Phi_{\rm TL}^2}[A^*] <
{\rm var}_{\Phi^2}[A^*]$ to leading order.

The thermal lines method combines the construction of a
one-dimensional two-point integration grid with $n$-dimensional Monte
Carlo sampling, and it is useful to analyze these two
aspects of the method separately.
If stochastic sampling is ignored, thermal lines reduces to quadrature
integration; in fact, the thermal lines grid for the harmonic wave
function is a two-point Gauss-Hermite quadrature grid.
\cite{Abramowitz_1965}
In the following, we generalize thermal lines by formulating
one-dimensional quadrature grids adapted to the single-mode wave
function in Section \ref{sec:1d_quadrature}, and then we separately
discuss the $n$-dimensional Monte Carlo sampling of these grids in
Section \ref{sec:mc_quadrature}.

\section{One-dimensional quadrature grids}
\label{sec:1d_quadrature}

We define our quadrature grids as an approximation to the integral
\begin{equation}
  \label{eq:quadrature_def}
  \langle A \rangle_{\phi^2} =
  \int_{-\infty}^\infty \phi^2(u) A(u) \, {\rm d}u \approx
    \sum_{\alpha=1}^p P_\alpha A(U_\alpha) \;,
\end{equation}
where $p$ is the number of points in the grid, $U_\alpha$ is the
$\alpha$th grid point, and $P_\alpha$ is its corresponding weight,
which satisfies $\sum_{\alpha=1}^p P_\alpha=1$.
This approximation is equivalent to replacing $\phi^2(u)$ in Eq.\@
\ref{eq:quadrature_def} with the discrete probability distribution
\begin{equation}
  \label{eq:grid_wfn_one_mode}
  \psi_p^2(u)
    = \sum_{\alpha=1}^p P_\alpha
                        \delta(u-U_\alpha) \;,
\end{equation}
so that the right-hand side of Eq.\@ \ref{eq:quadrature_def} is
$\langle A \rangle_{\psi_p^2}$.
Note that when $p\to\infty$ Eq.\ \ref{eq:grid_wfn_one_mode} must
reduce to the full single-mode probability distribution,
$\lim_{p\to\infty} P_\alpha \propto \phi^2(U_\alpha)$.

A $p$-point quadrature grid has $2p$ unknowns, which we determine
by imposing that the quadrature integral be exact for a polynomial
of order $2p-1$.
Let
\begin{equation}
  \label{eq:1D_A_local}
  A(u) = \sum_{\beta=0}^{2p-1} a_\beta u^\beta \;.
\end{equation}
The left-hand side of Eq.\@ \ref{eq:quadrature_def} is then
\begin{equation}
  \label{eq:1D_expval}
  \langle A \rangle_{\phi^2}
    = \sum_{\beta=0}^{2p-1} a_\beta \langle u^\beta \rangle_{\phi^2}
    = \sum_{\beta=0}^{2p-1} a_\beta \mu_\beta \;,
\end{equation}
where $\mu_\beta = \langle u^\beta \rangle_{\phi^2}$ is the $\beta$th
moment of $\phi^2$, and the right-hand side of Eq.\@
\ref{eq:quadrature_def} is
\begin{equation}
  \label{eq:1D_expval_grid}
  \langle A \rangle_{\psi_p^2}
    = \sum_{\beta=0}^{2p-1} a_\beta \langle u^\beta \rangle_{\psi_p^2}
    = \sum_{\beta=0}^{2p-1} a_\beta \sum_{\alpha=1}^p
                                    P_\alpha U_\alpha^\beta \;.
\end{equation}
Equating each term in the right-hand sides of Eqs.\@
\ref{eq:1D_expval} and \ref{eq:1D_expval_grid} yields
\begin{equation}
  \label{eq:grid_definition}
\left\{
  \sum_{\alpha=1}^p P_\alpha U_\alpha^\beta = \mu_\beta
\right\}_{\beta=0}^{2p-1} \;.
\end{equation}
The condition that the weights be normalized corresponds to $\beta=0$
in Eq.\@ \ref{eq:grid_definition}.
Note that, despite being derived for polynomial integrands,
quadrature grid integration can also accurately approximate integrals
of functions whose Taylor expansions do not converge in the
integration range, as demonstrated below.

Non-symmetric wave functions require solving the full system of
equations specified by Eq.\@ \ref{eq:grid_definition}.
Grids for symmetric wave functions must be symmetric, so if $U$ is a
grid point with weight $P$, then $-U$ must also be a grid point with
weight $P$, and consequently if $p$ is odd then $U=0$ must be a grid
point.
This eliminates $p$ equations from Eq.\@ \ref{eq:grid_definition} and
determines $p$ unknowns.
It is thus possible to obtain explicit analytical expressions for the
grid parameters of denser grids for symmetric wave functions than for
non-symmetric wave functions.
\begin{table}[hbt!]
  \begin{tabular}{ccccc}
    \hline \hline
    $\phi^2$ & $p$ & $\alpha$ & $U_\alpha$ & $P_\alpha$ \\
    \hline
    Symmetric     & 2 & 1 & $-\sqrt{\mu_2}$ & $1/2$ \\
                  &   & 2 & $ \sqrt{\mu_2}$ & $1/2$ \\[0.25cm]
                  & 3 & 1 & $-\sqrt{\frac{\mu_4} {\mu_2}}$
                          & $\frac{\mu_2^2} {2\mu_4}$ \\[0.1cm]
                  &   & 2 & $0$ & $1-\frac{\mu_2^2} {\mu_4}$ \\[0.1cm]
                  &   & 3 & $\sqrt{\frac{\mu_4} {\mu_2}}$
                          & $\frac{\mu_2^2} {2\mu_4}$ \\[0.25cm]
                  & 4 & 1 & $-\sqrt{\frac {\xi_6+\zeta_6} {2\xi_4} }$
                          & $\frac 1 4 -
                             \frac {\xi_6-2\mu_2 \xi_4}
                                   {4\zeta_6}$ \\[0.2cm]
                  &   & 2 & $-\sqrt{\frac {\xi_6-\zeta_6} {2\xi_4} }$
                          & $\frac 1 4 +
                             \frac {\xi_6-2\mu_2 \xi_4}
                                   {4\zeta_6}$ \\[0.2cm]
                  &   & 3 & $\sqrt{\frac {\xi_6-\zeta_6} {2\xi_4} }$
                          & $\frac 1 4 +
                             \frac {\xi_6-2\mu_2 \xi_4}
                                   {4\zeta_6}$ \\[0.2cm]
                  &   & 4 & $\sqrt{\frac {\xi_6+\zeta_6} {2\xi_4} }$
                          & $\frac 1 4 -
                             \frac {\xi_6-2\mu_2 \xi_4}
                                   {4\zeta_6}$ \\[0.2cm]
\multicolumn{2}{r}{} &
\multicolumn{3}{l}{where:} \\[0cm]
\multicolumn{3}{r}{} &
\multicolumn{2}{l}{$\xi_6 = \mu_6 - \mu_2\mu_4$} \\[0.1cm]
\multicolumn{3}{r}{} &
\multicolumn{2}{l}{$\xi_4 = \mu_4-\mu_2^2$} \\[0.1cm]
\multicolumn{3}{r}{} &
\multicolumn{2}{l}{$\zeta_6 = \sqrt{ \xi_6^2
                 - 4\xi_4 \left( \xi_6\mu_2 - \xi_4 \mu_4 \right)}$}
                            \\[0.3cm]
    Non-symmetric & 2 & 1 & $-\frac{2\mu_2^2}
                                   {\mu_3+\sqrt{\mu_3^2 + 4\mu_2^3}}$
                          & $\frac 1 2 +
                             \frac {\mu_3}
                                   {2\sqrt{\mu_3^2 + 4\mu_2^3}}$
                            \\[0.25cm]
                  &   & 2 & $\frac{\mu_3+\sqrt{\mu_3^2 + 4\mu_2^3}}
                                  {2\mu_2}$
                          & $\frac 1 2 -
                             \frac {\mu_3}
                                   {2\sqrt{\mu_3^2 + 4\mu_2^3}}$
                            \\[0.2cm]
    \hline \hline
  \end{tabular}
  \caption{
    \label{table:analytical_grids}
    Analytical expressions for the grid parameters obtained by solving
    Eq.\@ \ref{eq:grid_definition} for various grid sizes $p$ for
    symmetric and non-symmetric wave functions.
    Grid parameters for the harmonic wave function can be obtained by
    substituting $\mu_2=1/(2\omega)$, $\mu_4=3\mu_2^2$, and
    $\mu_6=15\mu_2^3$ in the expressions for symmetric grid
    parameters.
    Note that we assume $\mu_1=0$, which can always be accomplished
    by working with the shifted coordinate $u^\prime = u-\mu_1$;
    therefore in this table $\mu_\beta$ refers to the $\beta$th
    \textit{central} moment of $\phi^2(u)$.}
\end{table}
In Table \ref{table:analytical_grids} we give the parameters for
symmetric 2-, 3- and 4-point grids, and for the non-symmetric 2-point
grid.
Note that the symmetric 2-point grid corresponds to thermal lines,
and the non-symmetric 2-point grid reduces to the symmetric 2-point
grid if $\phi^2(u)$ has zero skewness.

\begin{figure}[htb!]
  \centering
  \includegraphics[width=0.48\textwidth]{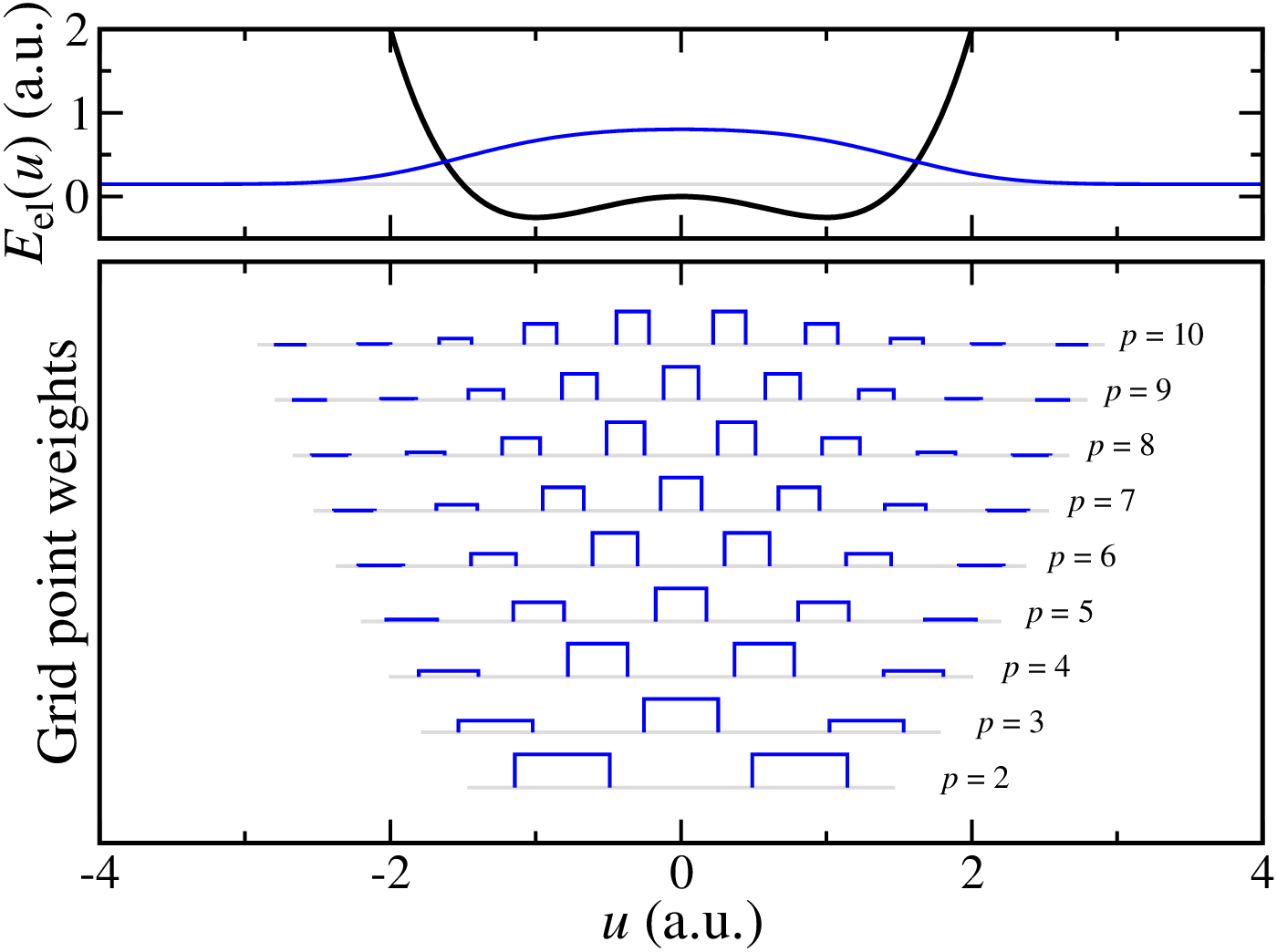}
  \caption{
    Symmetric anharmonic potential $E_{\rm el}(u) = -\frac 1 2 u^2 +
    \frac 1 4 u^4$ with its ground-state wave function (top panel),
    and quadrature grids constructed using this wave function (bottom
    panel).
    \label{fig:num_sym_grid}}
\end{figure}

\begin{figure}[htb!]
  \centering
  \includegraphics[width=0.48\textwidth]{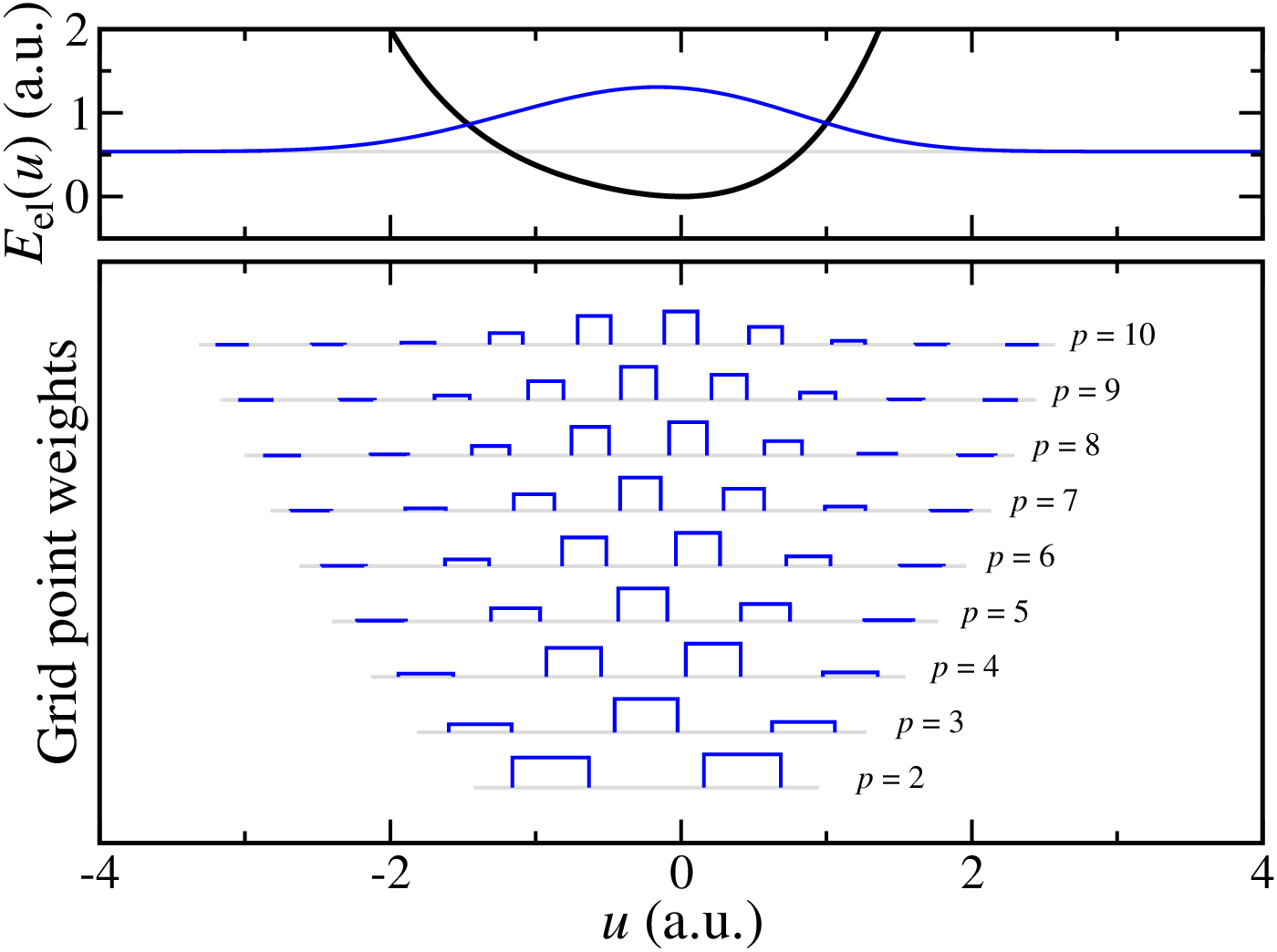}
  \caption{
    Non-symmetric anharmonic potential $E_{\rm el}(u) = \frac 1 2 u^2
    + \frac 1 4 u^3 + \frac 1 8 u^4$ with its ground-state wave
    function (top panel), and quadrature grids constructed using this
    wave function (bottom panel).
    \label{fig:num_nonsym_grid}}
\end{figure}

Larger grids can be constructed for any vibrational wave function by
numerically solving Eq.\@ \ref{eq:grid_definition}.
In Figs.\@ \ref{fig:num_sym_grid} and \ref{fig:num_nonsym_grid} we
show 2- to 10-point grids obtained for a symmetric and a non-symmetric
anharmonic potential, respectively.
The convergence of the grid weights to the square of the vibrational
wave function can be appreciated in the plots.

\begin{figure}[htb!]
  \centering
  \includegraphics[width=0.48\textwidth]{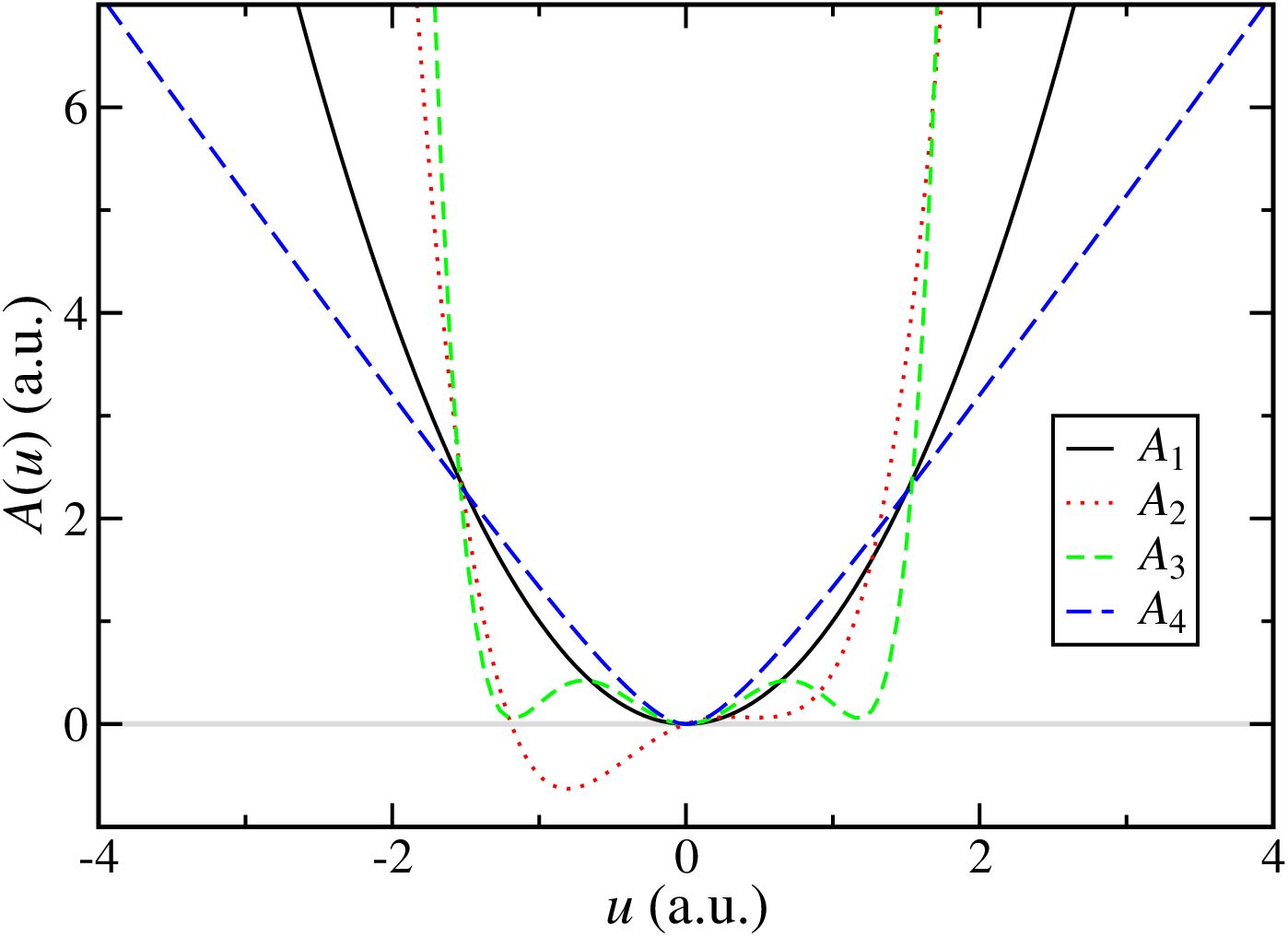}
  \caption{
    Test functions
    $A_1 = u^2$,
    $A_2 = 0.5u-u^2+u^4$,
    $A_3 = 2u^2-2.8u^4+u^6$, and
    $A_4 = \frac{4u^2}{1+2|u|}$,
    which serve as models of the dependence of an electronic
    observable $A$ on mode amplitude $u$.
    \label{fig:A_functions}}
\end{figure}

We test these numerical grids by integrating four test functions which
can be regarded as models of the dependence of an electronic
observable $A$ on the mode amplitude $u$.
Our test functions are a pure quadratic function, $A_1 = u^2$, a
non-monotonic, non-symmetric quartic polynomial which takes negative
values, $A_2 = 0.5 u-u^2+u^4$, a non-monotonic, symmetric,
non-negative sixth-order polynomial, $A_3 = 2u^2-2.8u^4+u^6$, and a
monotonic, non-negative, non-polynomial function locally dominated by
a quadratic term, $A_4 = \frac{4u^2}{1+2|u|}$.
These functions are plotted in Fig.\@ \ref{fig:A_functions}.
Functions similar to $A_2$ and $A_3$ have been used to model band gaps
in some systems, \cite{Whalley_perovskites_2016} while functions with
the linear asymptotic behavior of $A_4$ have been reported in
previous studies. \cite{monserrat_giant_2015,antonius_anh_2015}

The convergence of the variance of $A$ is particularly relevant to the
multi-dimensional Monte Carlo sampling of quadrature grids.
By construction, the bias in any expectation value obtained with a
$p$-point quadrature grid is ${\cal O}(u^{2p})$.
Therefore, if $A$ is a polynomial of order $n_A$, its quadrature grid
expectation value converges to the exact value at order
$p=\lceil \frac{n_A+1} 2 \rceil$, while the variance of $A$ is a
polynomial of order $2n_A$ and its quadrature grid estimate converges
to the exact variance at $p = n_A+1$.
Thus, if the variance approaches its exact value monotonically from
below, it would be possible to obtain the exact expectation value of
$A$ with a smaller uncertainty than with direct Monte Carlo by
stochastically sampling a quadrature grid of $\lceil \frac{n_A+1} 2
\rceil \leq p < n_A+1$ points.
Knowledge of $n_A$ for any given expectation value would then allow
selecting the grid size $p=\lceil \frac{n_A+1} 2 \rceil$ that
maximizes the efficiency of the stochastic sampling and incurs zero
bias.

\begin{figure}[htb!]
  \centering
  \includegraphics[width=0.48\textwidth]{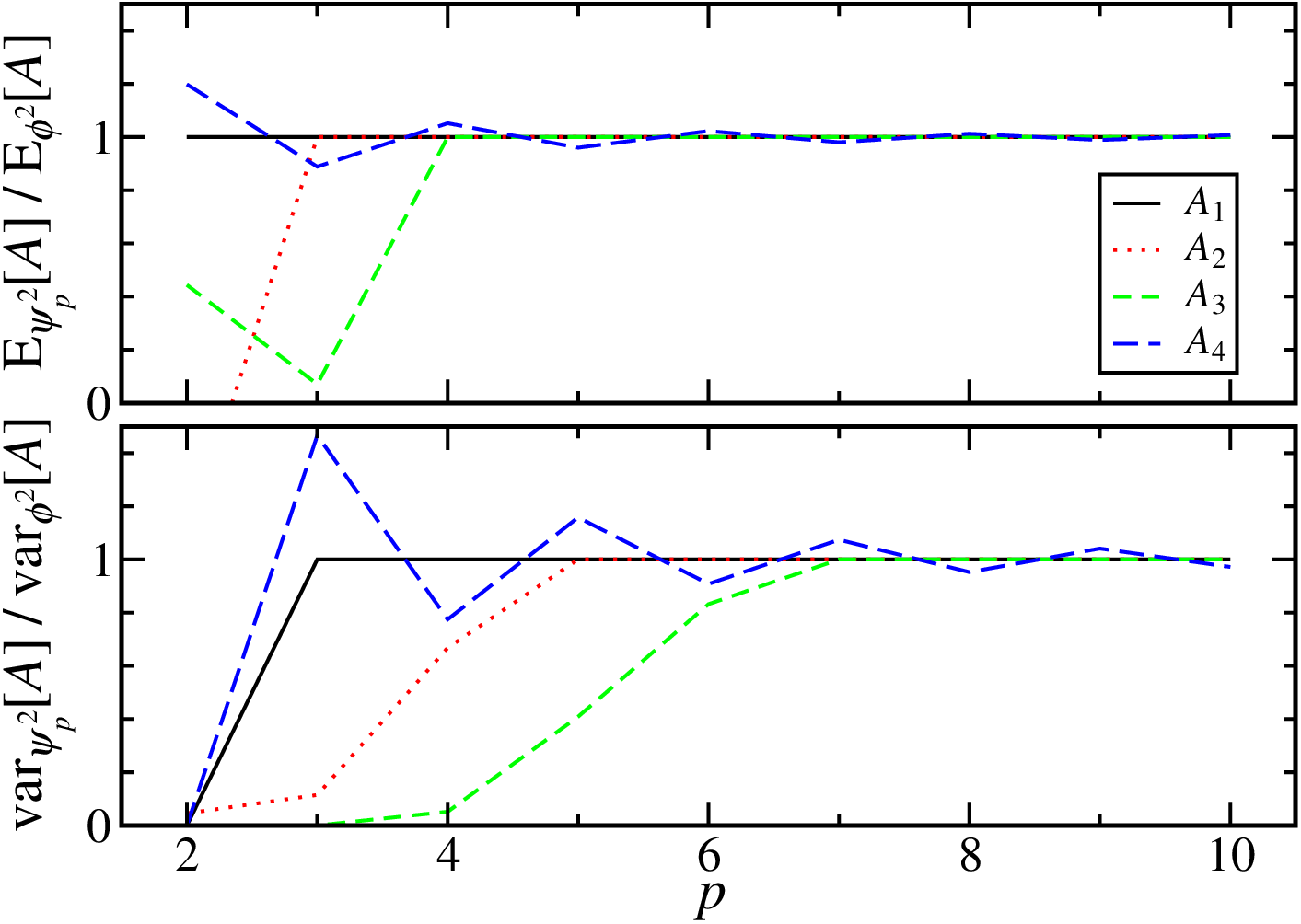}
  \caption{
    Quadrature-grid expectation value (top panel) and variance
    (bottom panel) of each of the four test functions plotted in
    Fig.\@ \ref{fig:A_functions} relative to the exact expectation
    value and variance, respectively, as a function of the number of
    points $p$ in the symmetric grids plotted in Fig.\@
    \ref{fig:num_sym_grid}.
    \label{fig:num_sym_results}}
\end{figure}

\begin{figure}[htb!]
  \centering
  \includegraphics[width=0.48\textwidth]{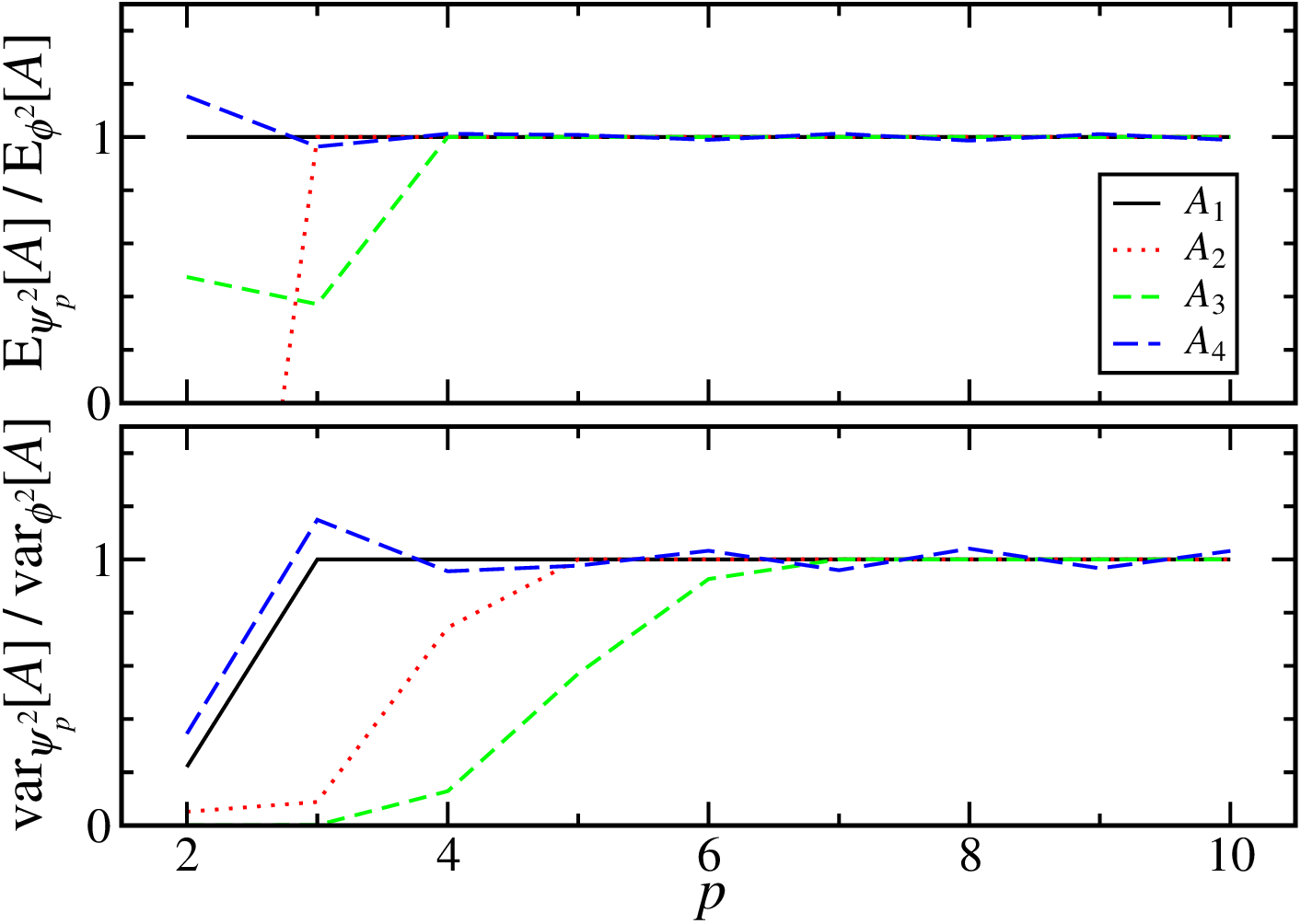}
  \caption{
    Quadrature-grid expectation value (top panel) and variance
    (bottom panel) of each of the four test functions plotted in
    Fig.\@ \ref{fig:A_functions} relative to the exact expectation
    value and variance, respectively, as a function of the number of
    points $p$ in the non-symmetric grids plotted in Fig.\@
    \ref{fig:num_nonsym_grid}.
    \label{fig:num_nonsym_results}}
\end{figure}

In Figs.\@ \ref{fig:num_sym_results} and \ref{fig:num_nonsym_results}
we plot the expectation value and variance of the four test functions
for the symmetric and non-symmetric numerical grids, respectively.
In these tests, the integrals of polynomials of order $n_A$ are
approximated particularly poorly by grids with $p<\lceil
\frac{n_A+1} 2 \rceil$ points.
For $\lceil \frac{n_A+1} 2 \rceil \leq p < n_A+1$ the quadrature
integrals are exact, as expected, while the variance approaches
its exact value from below, making stochastic quadrature grid
integration advantageous over direct Monte Carlo at these grid sizes.

The expectation value and variance of the non-polynomial $A_4$ test
function converge slowly and non-monotonically to their respective
exact values, but the quadrature integrals of $A_4$ are reasonably
good approximations to the exact integral at all $p$, and two-point
grids are particularly efficient since the corresponding variances
are small (or indeed, zero for a symmetric wave function).

\section{$n$-dimensional Monte Carlo sampling of one-dimensional
quadrature grids}
\label{sec:mc_quadrature}

We define quadrature grids for each of the $n$ normal-mode
coordinates, which are stochastically sampled by choosing random grid
points with probability proportional to their weights.
The basic properties of expectation values of functions of $n$
variables differ from those in one dimension due to the effect of
multi-mode contributions.
As in the case of thermal lines, $n$-dimensional sampling of $p$-point
quadrature grids correctly accounts for multi-mode contributions to
the expectation value involving up to the $(2p-1)$th power of
normal-mode coordinates, exceeding the order to which single-mode
contributions are exact.
Therefore we do not expect major differences in the behavior and
convergence properties of $n$-dimensional quadrature grid expectation
values with respect to the one-dimensional case.

Observables along most of the normal modes in a system are found in
practice to be strongly quadratic, and therefore it is particularly
interesting to investigate the use of per-mode grid sizes adapted to
the specific system under consideration.
We test $n$-dimensional sampling of quadrature grids by evaluating
the zero-point correction to the band gap of a primitive cell of the
HF and NH$_3$ molecular crystals using the harmonic vibrational wave
function.
These systems were found to be particularly problematic under the
quadratic approximation in Ref.\@ \onlinecite{monserrat_giant_2015},
and therefore our results should be representative of the usefulness
of multi-point quadrature grids in difficult cases.
The band gap of HF is a markedly non-quadratic function of the
highest-frequency normal mode, similar in shape to the one-dimensional
test function $A_4$ in Section \ref{sec:1d_quadrature}, see Fig.\@ 2
of Ref.\@ \onlinecite{monserrat_giant_2015}, while it is found to be a
strongly quadratic function of the 20 remaining normal modes.
We similarly find that the non-quadratic behavior of the band gap of
NH$_3$ arises largely from two mid-frequency normal modes, with the 43
remaining modes providing mostly quadratic contributions.
This information allows us to choose which normal modes to sample
using grids of $p>2$ points.

The local value of the band gap at each nuclear configuration is
evaluated using the plane-wave DFT method with the PBE functional
\cite{perdew_pbe_1996} as implemented in the \textsc{castep} code,
\cite{clark_castep_2005} and vibrational calculations are performed
using our own code.
We use symmetrized sampling for all of our calculations, and we
test (i) varying the number of grid points along all normal modes,
and (ii) varying the number of grid points only along the modes for
which we find the band gap to be non-quadratic, using two-point
grids for the remaining modes.
\begin{figure}[htb!]
  \centering
  \includegraphics[width=0.48\textwidth]{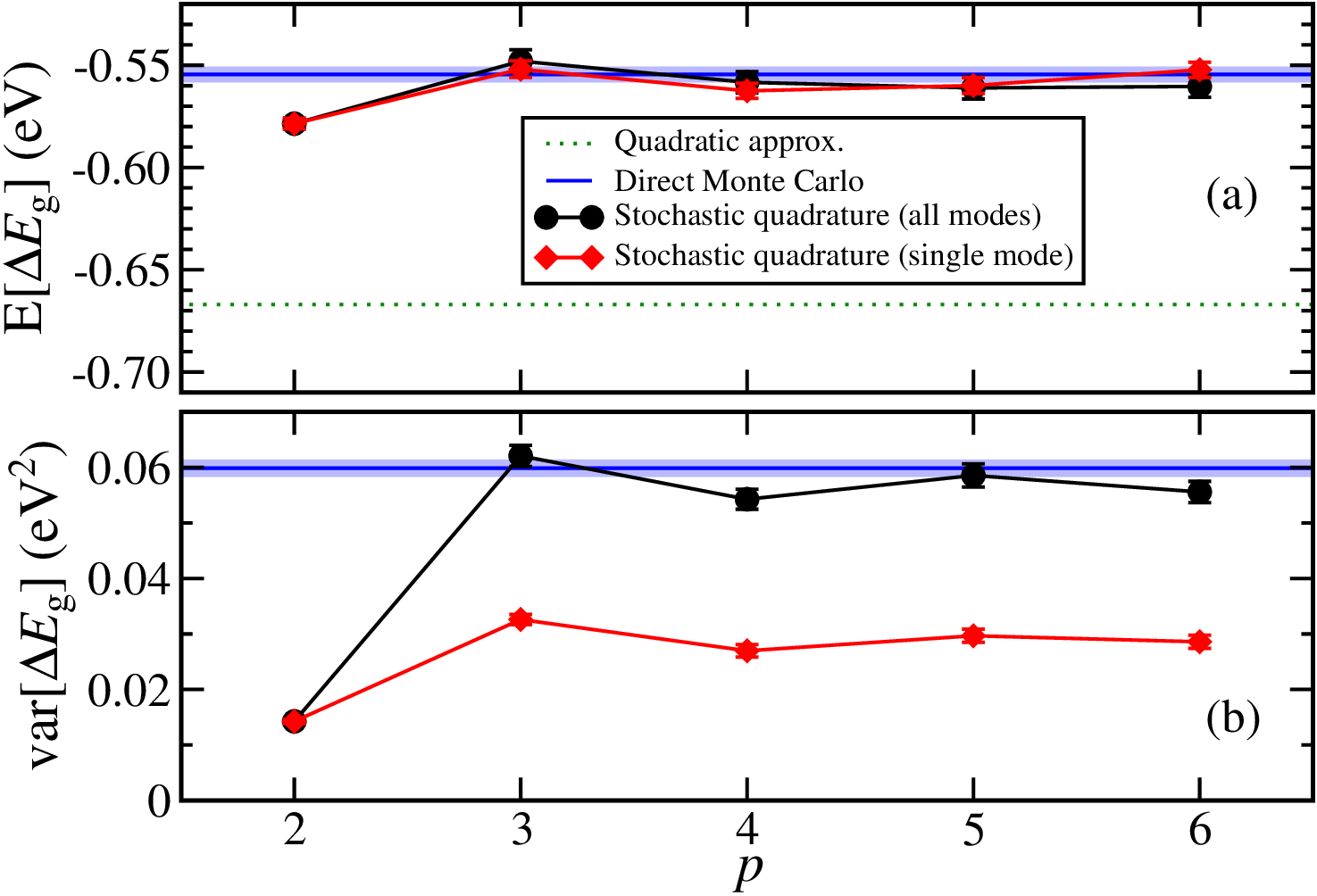}
  \caption{
    (a) expectation value and (b) sample variance of the zero-point
    correction to the band gap of a primitive cell of the HF
    molecular crystal as a function of the grid size along all
    normal modes (black circles) and only along the mode for which
    the band gap is found to be non-quadratic, with the remaining
    modes using two-point grids (red diamonds).
    The horizontal blue lines represent the exact results evaluated by
    direct Monte Carlo integration, the light-colored areas represent
    their standard errors, and the horizontal green dotted line is the
    expectation value obtained with the quadratic approximation.
    \label{fig:hf_gap}}
\end{figure}
\begin{figure}[htb!]
  \centering
  \includegraphics[width=0.48\textwidth]{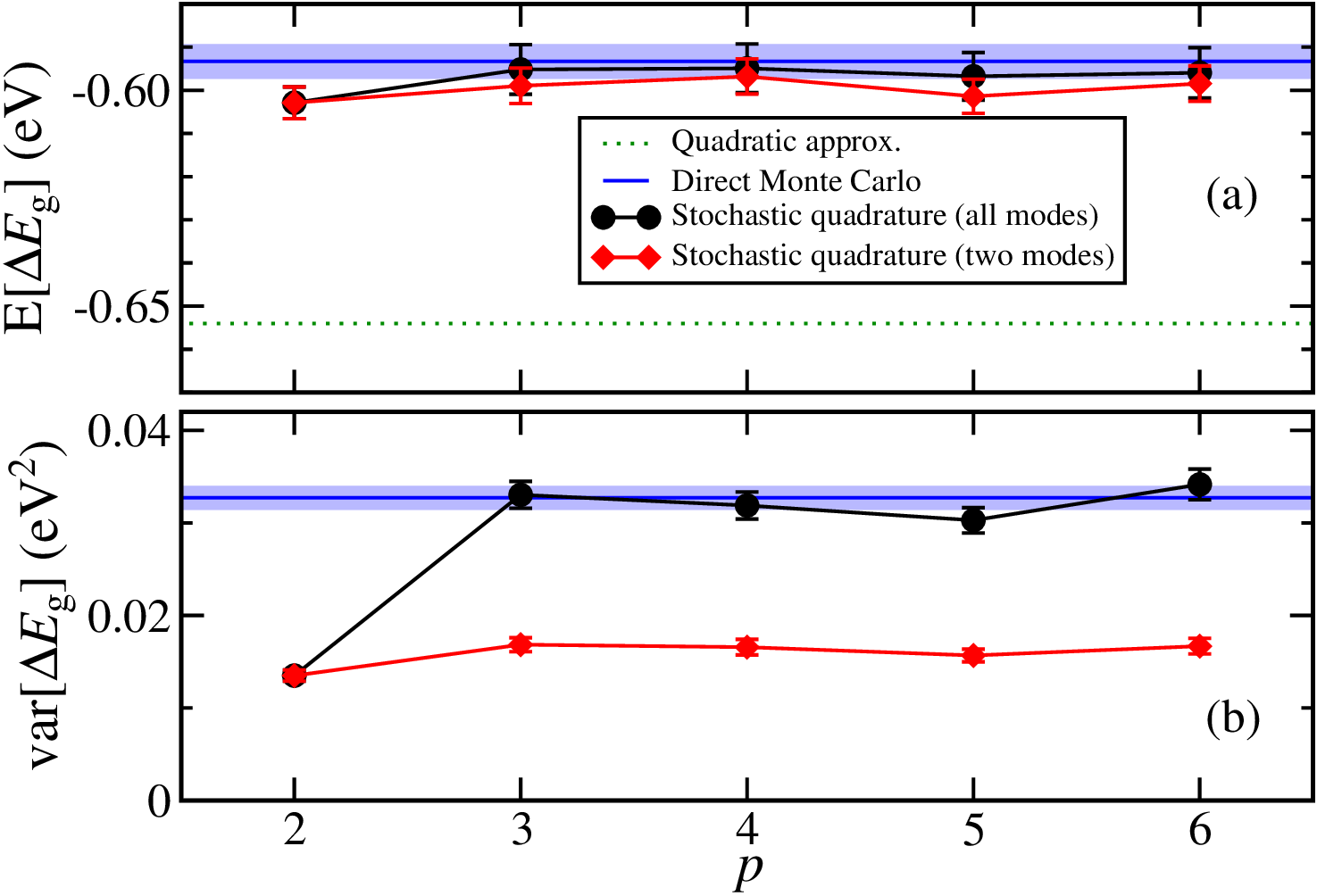}
  \caption{
    (a) expectation value and (b) sample variance of the zero-point
    correction to the band gap of a primitive cell of the NH$_3$
    molecular crystal as a function of the grid size along all
    normal modes (black circles) and only along the two modes for
    which the band gap is found to be non-quadratic, with the
    remaining modes using two-point grids (red diamonds).
    The horizontal blue lines represent the exact results evaluated by
    direct Monte Carlo integration, the light-colored areas represent
    their standard errors, and the horizontal green dotted line is the
    expectation value obtained with the quadratic approximation.
    \label{fig:nh3_gap}}
\end{figure}
The expectation value and variance of the band gap of HF and NH$_3$
are plotted in Figs.\@ \ref{fig:hf_gap} and \ref{fig:nh3_gap},
respectively, as a function of the number of points $p$ in
the integration grid.

The results using $p$ grid points along all normal modes converge
quickly towards the infinite grid limit.
The sample variance is within statistical uncertainty of this limit
for $p\geq 3$, implying that, while quadrature grids with $p>2$ offer
excellent results for these systems, they offer no speed advantage
over direct Monte Carlo integration.
Only two-point grids (thermal lines) provide a significant reduction
in statistical uncertainty, while still improving upon the quadratic
approximation thanks to correctly accounting for multi-mode
contributions.

In the case of the HF molecular crystal, limiting the use of grids
with $p>2$ points to the normal mode of which the band gap is a
non-quadratic function gives expectation values within statistical
uncertainty of the direct Monte Carlo result, but with about half
the sample variance.
This implies a $\sim 30\%$ reduction in the number of samples needed
to obtain the expectation value to a target statistical uncertainty.
For NH$_3$ the uncertainty in the expectation value of the band gap
prevents drawing strong conclusions regarding the selective
application of multi-point grids, but the results hint at a reduced
bias compared with $p=2$, and the sample variance is again about half
that obtained with direct Monte Carlo.

These results reflect that, even when non-quadratic behavior is
present in an observable, the overall weight of non-quadratic
contributions to the result is typically small; two-point grids
provide a very good approximation for quadratic contributions, and
multi-point grids can be used along specific modes to avoid any bias.

\section{Conclusions}
\label{sec:conclusions}

We have quantified the bias incurred by the thermal lines method,
which includes higher-order terms than the quadratic approximation,
and we have reformulated thermal lines as a particular choice of grid
size in a stochastic implementation of quadrature grid integration.
We have demonstrated the construction of one-dimensional quadrature
grids adapted to specific anharmonic wave functions and their
application to the integration of functions modelling the dependence
of electronic observables on nuclear configurations.

The accuracy of quadrature integration ultimately depends on the
function being integrated.
Our tests with model polynomials show that small grid sizes can incur
a large bias, but knowledge of the order of the polynomial allows
exploiting the slow convergence of the variance with grid size to
obtain accurate expectation values with little statistical noise.

Our tests using DFT data indicate that observables of interest
tend to be dominated by quadratic contributions, for which two-point
grids (thermal lines) are ideally suited.
However examples have been reported of observables with strong
non-quadratic components along specific modes,
\cite{monserrat_giant_2015, antonius_anh_2015,
Whalley_perovskites_2016} and we find that selectively using
quadrature grids with three or more points along these modes
eliminates the bias in the expectation values.

We therefore recommend that the behavior of observables along each
normal mode be determined prior to the application of stochastic
quadrature grid integration in order to select the optimal grid size
and avoid incurring a bias.
This type of analysis can be trivially performed during the
mapping of the Born-Oppenheimer energy surface $E_{\rm el}({\bf u})$
involved in solving anharmonic vibrational Hamiltonians, and can
be carried out separately by direct inspection when using the harmonic
approximation.

\begin{acknowledgments}
  P.L.R.\@  acknowledges financial support from the Max-Planck
  Society.
  R.J.N.\@ acknowledges financial support from the
  Engineering and Physical Sciences Research Council, U.K., under
  grant no.\@ EP/P034616/1.
  B.M.\@ acknowledges Robinson College, Cambridge, and the Cambridge
  Philosophical Society for a Henslow Research Fellowship.
  We are grateful for computational support from the UK national high
  performance computing service, ARCHER, for which access was obtained
  via the UKCP consortium and funded by EPSRC grant no.\@
  EP/P022561/1.
  Supporting research data may be freely accessed at [URL], in
  compliance with the applicable Open Data policies.
\end{acknowledgments}


\end{document}